\documentstyle[aps,amssymb,psfig]{revtex}
\begin{document}

\title{Kadanoff--Baym equations and non-Markovian Boltzmann equation in
generalized T--matrix approximation}
\author	{D. Semkat, D. Kremp, and M. Bonitz}
\address{Universit\"at Rostock, Fachbereich Physik, Universit\"atsplatz 3,
D--18051 Rostock }
\date{\today}
\maketitle

\begin{abstract}

A recently developed method \cite{SKB99,KSB-KB} for incorporating initial binary
correlations into the Kadanoff--Baym equations (KBE) is used to derive a
generalized T--matrix approximation for the self-energies. It is shown that the
T--matrix obtains additional contributions arising from initial correlations.
Using these results and taking the time-diagonal limit of the KBE, a generalized
quantum kinetic equation in binary collision approximation is derived.
This equation is a far-reaching generalization of Boltzmann--type kinetic
equations: it selfconsistently includes memory effects (retardation, off-shell
T--matrices) as well as many-particle effects (damping, in-medium
T--matrices) and spin-statistics effects (Pauli--blocking).

\end{abstract}


\pacs{05.20.Dd, 52.25.Dg}

\section{Introduction}\label{intro}

Nonequilibrium properties of many-particle systems have traditionally been
described by kinetic equations of the Boltzmann type. Despite their fundamental
character, these equations have well-known principal shortcomings, e.g.
(i) the short-time behavior ($t<\tau_{cor}$ - the correlation time) cannot be
described correctly, (ii) the kinetic or the quasiparticle energy is conserved
instead of the total (sum of kinetic and potential) energy, (iii) no bound
states are contained, and (iv) in the long-time limit, they yield the
equilibrium distribution and thermodynamics of ideal particles.

An important generalization are the well-known Kadanoff--Baym equations
derived by Kadanoff and Baym \cite{KadBaym}, and Keldysh \cite{Keldysh}.
However, the original KBE contain no contribution from initial correlations.
Therefore, the KBE are unable to describe the initial stage of the evolution
($t_0 \le t \le \tau_{cor}$) and the influence of initial correlations which
can be important for ultrafast relaxation processes.

To include initial correlations into the KBE, various methods have been used,
including analytical continuation of the equilibrium KBE to real times
\cite{KadBaym,Dan84_1,Wagner,MR,MR-KB} and perturbation theory
with initial correlations \cite{Fujita,Hall,Dan84_1}.
A convincing solution has been presented by Danielewicz \cite{Dan84_1}, who
developed a perturbation theory for a general initial state and derived
generalized KBE which take into account arbitrary initial correlations.
Finally, a straightforward and very intuitive method which does not make use of
perturbation theory but uses the equations of motion for the Green's functions
instead, has been developed in \cite{SKB99,KSB-KB}. While perturbative
approaches are restricted to situations where the coupling is weak, our method
is valid for arbitrary coupling strength. In particular, it allows to
consider systems with strong coupling, such as Coulomb systems at low
temperatures and/or high density (e.g. metals and dense plasmas) and nuclear 
matter, and to include bound states.
In Sec.~\ref{ic} we briefly recall the main ideas of our method. After this,
Sec.~\ref{tmat} is devoted to the application of our approach to the T--matrix
approximation. In Sec.~\ref{Boltzmann} we derive a non-Markovian Boltzmann
equation in binary collision approximation.

\section{Initial correlations in the Kadanoff--Baym equations}\label{ic}

Starting point of our approach is the first equation of the Martin--Schwinger
hierarchy \cite{note1},
\begin{eqnarray}\label{MSH}
(S_{ac}-U_{ac})G_{cb}&=&\delta_{ab}\pm i\hbar\,V_{ad,ce}G_{ce,bd},\\
\mbox{with}\quad
S_{ac}&=&\left(i\hbar\frac{\partial}{\partial t_a}
+\frac{\hbar^2\nabla_a^2}{2m_a}\right)\delta_{ac},
\end{eqnarray}
together with an initial condition for $G_{ce,bd}$,
\begin{eqnarray}\label{AB}
G_{ce,bd}|_{t_c=t_e=t_b=t_d=t_0}=
G_{cb}(t_0)G_{ed}(t_0)\pm G_{cd}(t_0)G_{eb}(t_0)+C_{ce,bd}(t_0).
\end{eqnarray}
Summation/integration over repeated indices is implied.
Here, $C$ denotes initial binary correlations in the system, and $U$ is an
external potential. The self-energy is defined by
\begin{eqnarray}\label{Sigmadef}
\Sigma_{ac}G_{cb}=\pm i\hbar\,V_{ad,ce}G_{ce,bd}
=\pm i\hbar\,V_{ad,ce}\left\{G_{cb}G_{ed}
\pm\frac{\delta G_{cb}}{\delta U_{de}}\right\}.
\end{eqnarray}
Considering Eq.~(\ref{Sigmadef}) in the limit $t=t'\to t_0$, we get explicitly
\begin{eqnarray}\label{Sigmat0}
\int d{\bar t}\,\Sigma_{ac}(t_0,{\bar t})G_{cb}({\bar t},t_0)
=\pm i\hbar\,V_{ad,ce}
\left\{G_{cb}(t_0)G_{ed}(t_0)\pm G_{cd}(t_0)G_{eb}(t_0)+C_{ce,bd}(t_0)\right\}.
\end{eqnarray}
Since the time integration is performed along the Keldysh--Schwinger contour,
only time-local contributions of $\Sigma$ survive on the l.h.s.
The last term on the r.h.s shows that there must exist, in addition to the
Hartree--Fock contributions (first two terms), another time-local
part, which is related to initial correlations. That means, the
self-energy has the structure (${\hat\Sigma}$ denotes the self-energy in the
adjoint equation)
\begin{eqnarray}\label{Sigmastruc}
\Sigma_{ab}&=&\Sigma^{HF}_{ab}+\Sigma^C_{ab}+\Sigma^{IN}_{ab},\\
{\hat\Sigma}_{ab}&=&\Sigma^{HF}_{ab}+\Sigma^C_{ab}+{\hat\Sigma}^{IN}_{ab},
\end{eqnarray}
with the time-local terms (here, we give the time arguments explicitly)
\begin{eqnarray}\label{Sigmainstruc}
\Sigma^{IN}_{ab}(t,t')&=&\Sigma^{IN}_{ab}(t,t_0)\delta(t_0-t'),\\
{\hat\Sigma}^{IN}_{ab}(t,t')&=&{\hat\Sigma}^{IN}_{ab}(t_0,t')\delta(t-t_0).
\end{eqnarray}

The further steps aim at the determination of these initial correlation terms
and are sketched here, for details, we refer to Refs.~\cite{SKB99,KSB-KB}.
Inserting (\ref{Sigmadef}) into (\ref{MSH}), one obtains a Dyson--Schwinger
equation for $t,t'>t_0$,
\begin{eqnarray}\label{Dyson}
(S_{ac}-U_{ac}-\Sigma_{ac})G_{cb}=\delta_{ab},
\end{eqnarray}
which can be cast into the form $G_{ac}^{-1}G_{cb}=\delta_{ab}$.
Functional differentiation of this equation with respect to the external
potential $U$ yields a Bethe--Salpeter equation for $\delta G/\delta U$.
Performing the same steps for the adjoint equation to (\ref{MSH}) as well,
a solution for $\delta G/\delta U$, which incorporates initial binary
correlations, is obtained,
\begin{eqnarray}\label{dgdu}
\frac{\delta G_{ab}}{\delta U_{dc}}=G_{ad}G_{cb}
+G_{ae}\frac{\delta
\left[\Sigma^C_{ef}+\Sigma^{IN}_{ef}+{\hat\Sigma}^{IN}_{ef}\right]}
{\delta U_{dc}}G_{fb}
\pm G_{ae}G_{cf}C_{ef,gh}G_{gb}G_{hd},
\end{eqnarray}
where $C$ has the time structure
\begin{eqnarray}\label{C}
C_{ab,cd}(t_at_b,t_ct_d)=C_{ab,cd}(t_0)\delta(t_a-t_0)\delta(t_b-t_0)
\delta(t_c-t_0)\delta(t_d-t_0).
\end{eqnarray}

\section{Generalized T--matrix approximation}\label{tmat}

In the previous section we have obtained a formal decoupling of the
Martin--Schwinger hierarchy by introduction of the self-energy. Furthermore,
our approach shows, that initial correlations can, in principle, be
straightforwardly included into this quantity. The next step on the way to a
quantum kinetic equation is to choose a suitable approximation for the
self-energy. Among the standard schemes are the random phase approximation
(RPA), describing dynamical screening, and the T--matrix (or binary collision)
approximation. The determination of $\Sigma$ in these schemes without inclusion
of initial correlations is well-known. For example, the T--matrix approximation
leads to a non-Markovian Boltzmann equation. In Ref.~\cite{KBKS97}, this
equation has been derived within the density operator technique.
The nonequilibrium Green's functions approach, however, opens the possibility
to derive two-time quantum kinetic equations with their well-known
advantages (e.g. they fully include the kinetic and spectral one-particle
properties). One-time equations are obtained by taking the time-diagonal limit
of the two-time equations in a much simpler way than within the density
operator technique.

In the following, we will use the nonequilibrium Green's functions theory to
derive a generalization of the usual T--matrix approximation, which includes
initial binary correlations.

According to Eqs.~(\ref{Sigmadef},\ref{dgdu}), the self-energy is determined
by the functional equations \cite{note2}
\begin{eqnarray}\label{Sigma}
\Sigma_{ab}&=&\pm i\hbar\,
V_{ad,ce}\left\{\delta_{cb}G_{ed}\pm \delta_{eb}G_{cd}
+G_{cf}G_{eg}C_{fg,bh}G_{hd}\pm G_{cf}
\frac{\delta\left[\Sigma_{fb}+{\hat\Sigma}^{IN}_{fb}\right]}
{\delta U_{de}}\right\}
,\\
\label{Sigmadach}
{\hat\Sigma}_{ab}&=&\pm i\hbar
\left\{\delta_{ae}G_{cd}\pm \delta_{ac}G_{de}
+G_{dg}C_{ag,fh}G_{fc}G_{he}
\pm\frac{\delta\left[{\hat\Sigma}_{af}+\Sigma^{IN}_{af}\right]}
{\delta U_{ed}}G_{fc}
\right\}V_{ce,bd}
,\\
\label{Sigma-in}
\Sigma^{IN}_{ab}&=&\pm i\hbar\,
V_{ad,ce}\left\{G_{cf}G_{eg}C_{fg,bh}G_{hd}\pm G_{cf}
\frac{\delta\Sigma^{IN}_{fb}}{\delta U_{de}}\right\},\\
\label{Sigma_in}
{\hat\Sigma}^{IN}_{ab}&=&\pm i\hbar
\left\{G_{dg}C_{ag,fh}G_{fc}G_{he}
\pm\frac{\delta{\hat\Sigma}^{IN}_{af}}
{\delta U_{ed}}G_{fc}
\right\}V_{ce,bd}.
\end{eqnarray}

Notice especially that, due to the structure of the self-energy, the
arguments of the functional derivative in the equations for $\Sigma$ and
${\hat\Sigma}$ are the same in both cases,
\begin{eqnarray}\label{tildeSigmadef}
\Sigma+{\hat\Sigma}^{IN}={\hat\Sigma}+\Sigma^{IN}
=\Sigma^C+\Sigma^{IN}+{\hat\Sigma}^{IN}={\tilde\Sigma}.
\end{eqnarray}

We now introduce an effective two-particle potential $\Xi$ by
\begin{eqnarray}\label{Xidef}
\frac{\delta{\tilde\Sigma}_{ab}}{\delta U_{cd}}=
\frac{\delta{\tilde\Sigma}_{ab}}{\delta G_{ef}}
\frac{\delta G_{ef}}{\delta U_{cd}}=
\pm i\hbar\,\Xi_{af,be}\frac{\delta G_{ef}}{\delta U_{cd}},
\end{eqnarray}
and define a generalized T--matrix \cite{BW63},
\begin{eqnarray}\label{Tdef}
{\cal T}_{ab,cd}=\Xi_{ab,cd}\pm i\hbar\,\Xi_{ae,cf}G_{fg}G_{he}{\cal T}_{gb,hd}
\pm\Xi_{ae,cf}G_{fg}G_{he}C_{gb,hd}.
\end{eqnarray}
In terms of Feynman diagrams, Eq.~(\ref{Tdef}) reads (the shaded block denotes
the initial correlation $C$)
\begin{eqnarray*}
\centerline{ 
\psfig{file=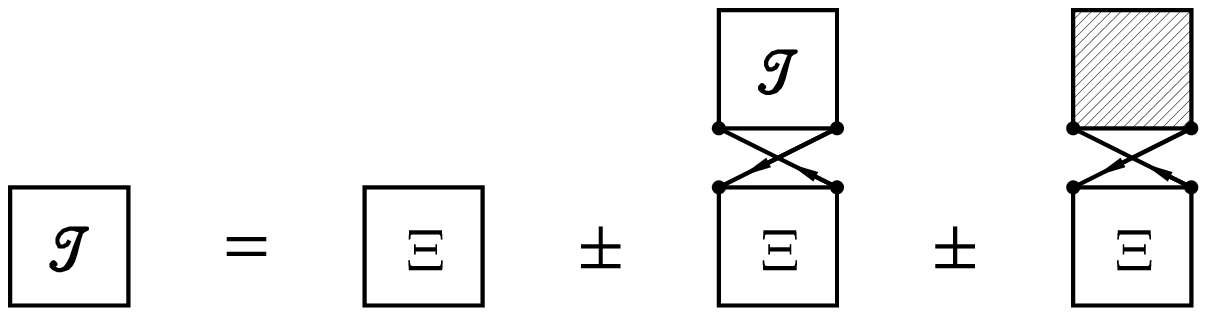,height=2.625cm,width=10.5cm}}
\end{eqnarray*}

Comparing Eq.~(\ref{Tdef}) with the solution (\ref{dgdu}) for
$\delta G/\delta U$, one obtains the relation
\begin{eqnarray}\label{T_Sig_U}
\frac{\delta{\tilde\Sigma}_{ab}}{\delta U_{cd}}=
\pm i\hbar\, G_{de}{\cal T}_{ae,bf}G_{fc}.
\end{eqnarray}
So we could identify ${\cal T}$ with the correlated part of the two-particle
function without the bare initial correlation $C$.
The equation for the self-energy now takes the form
\begin{eqnarray}\label{Sigma_T}
\Sigma_{ab}&=&\pm i\hbar\,
V_{ad,ce}\left\{\delta_{cb}G_{ed}\pm \delta_{eb}G_{cd}
+G_{cf}G_{eg}C_{fg,bh}G_{hd}
+i\hbar\,G_{cf}G_{eg}{\cal T}_{fg,bh}G_{hd}
\right\}.
\end{eqnarray}
Functional differentiation of this equation yields a relation for $\Xi$,
which depends on ${\cal T}$ and on the quantity
$\delta{\hat\Sigma}^{IN}/\delta G \equiv \pm i\hbar\,\Phi$. Inserting this
relation into Eq.~(\ref{Tdef}), and evaluating the functional derivative
$\delta{\hat\Sigma}^{IN}/\delta G$, one arrives at two coupled equations for
${\cal T}$ and $\Phi$, where self-energies and $\Xi$ have been eliminated.
Keeping only the ladder-type terms, these equations can be written as
\begin{eqnarray}\label{T_Leiter}
{\cal T}_{ab,cd}&=&V_{ab,cd}+\Phi_{ab,cd}+V_{ab,ef}G_{eg}G_{fh}C_{gh,cd}
+i\hbar\,V_{ab,ef}G_{eg}G_{fh}{\cal T}_{gh,cd},\\
\label{Phi_Leiter}
\Phi_{ab,cd}&=&C_{ab,ef}G_{eg}G_{fh}V_{gh,cd}
+i\hbar\,\Phi_{ab,ef}G_{eg}G_{fh}V_{gh,cd}.
\end{eqnarray}
Eqs.~(\ref{T_Leiter},\ref{Phi_Leiter}) can be solved easily (see Appendix A),
yielding an explicit expression for ${\cal T}$,
\begin{eqnarray}\label{T_expl}
{\cal T}_{ab,cd}=T_{ab,cd}
&+&i\hbar\, T_{ab,ef}G_{eg}G_{fh}C_{gh,ij}G_{ik}G_{jl}T_{kl,cd}
\nonumber\\
&+&T_{ab,ef}G_{eg}G_{fh}C_{gh,cd}
+C_{ab,ef}G_{eg}G_{fh}T_{gh,cd},
\end{eqnarray}
or, in terms of Feynman diagrams,
\begin{eqnarray*}
\centerline{
\psfig{file=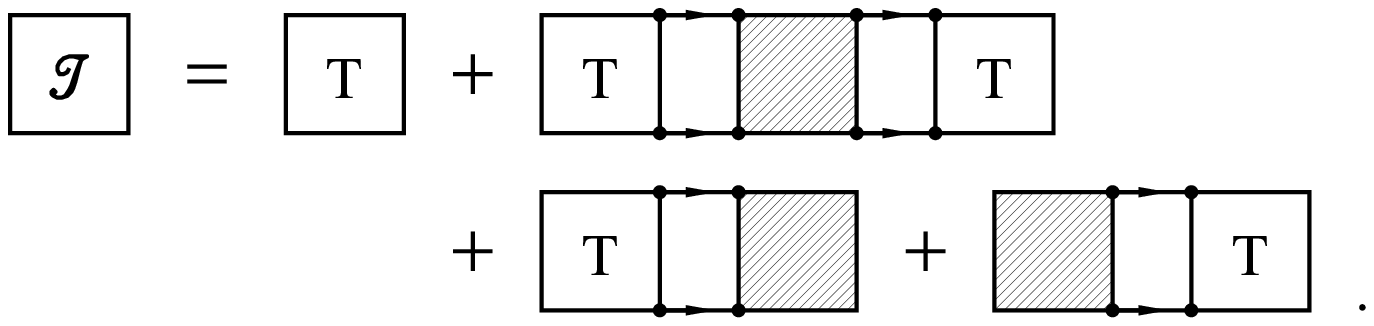,height=2.625cm,width=11.375cm}}
\end{eqnarray*}
Here, $T$ denotes the well-known ``ladder T--matrix'' which obeys
\begin{eqnarray}\label{T_LSG}
T_{ab,cd}=V_{ab,cd}+i\hbar\,V_{ab,ef}G_{eg}G_{fh}T_{gh,cd},
\end{eqnarray}
\begin{eqnarray*}
\centerline{
\psfig{file=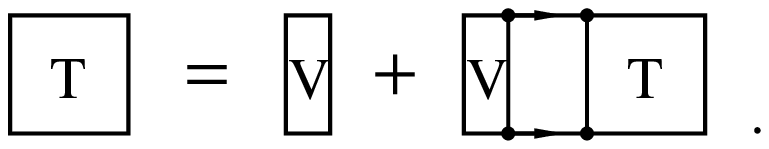,height=1.5cm,width=6.75cm}}
\end{eqnarray*}
The system (\ref{T_Leiter},\ref{Phi_Leiter}) can be regarded as a generalization
of the usual T--matrix equation (\ref{T_LSG}), where Eq.~(\ref{T_expl}) shows
explicitly the corrections which are due to initial correlations.

If we now insert Eq.~(\ref{T_expl}) into the equation for the self-energy
(\ref{Sigma_T}), we obtain $\Sigma$ in T--matrix (binary collision)
approximation,
\begin{eqnarray}\label{Sigma_T_bin}
\Sigma_{ac}=\pm i\hbar\, T_{ab,cd}G_{db}
&\pm& i\hbar\, T_{ab,ef}G_{eg}G_{fh}C_{gh,cd}G_{db}\nonumber\\
&\pm& (i\hbar)^2\, T_{ab,ef}G_{eg}G_{fh}C_{gh,ij}G_{ik}G_{jl}T_{kl,cd}G_{db},
\end{eqnarray}
\begin{eqnarray*}
\centerline{
\psfig{file=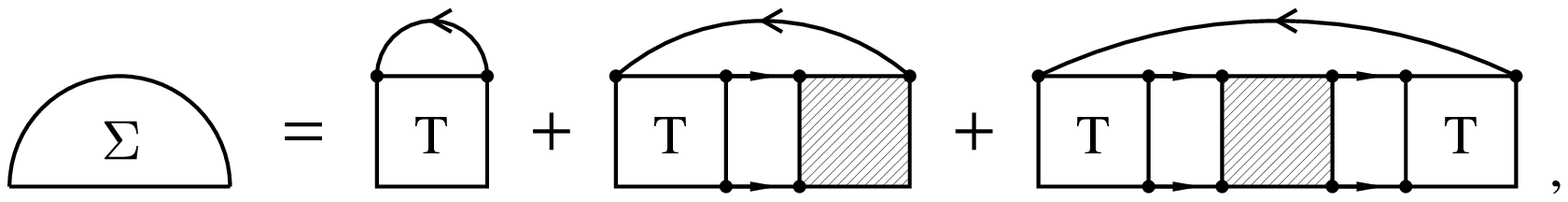,height=1.875cm,width=13.125cm}}
\end{eqnarray*}
and analogously ${\hat\Sigma}$,
\begin{eqnarray}\label{Sigmadach_T_bin}
{\hat\Sigma}_{ac}=\pm i\hbar\, T_{ab,cd}G_{db}
&\pm& i\hbar\, C_{ab,ef}G_{eg}G_{fh}T_{gh,cd}G_{db}\nonumber\\
&\pm& (i\hbar)^2\, T_{ab,ef}G_{eg}G_{fh}C_{gh,ij}G_{ik}G_{jl}T_{kl,cd}G_{db},
\end{eqnarray}
\begin{eqnarray*}
\centerline{
\psfig{file=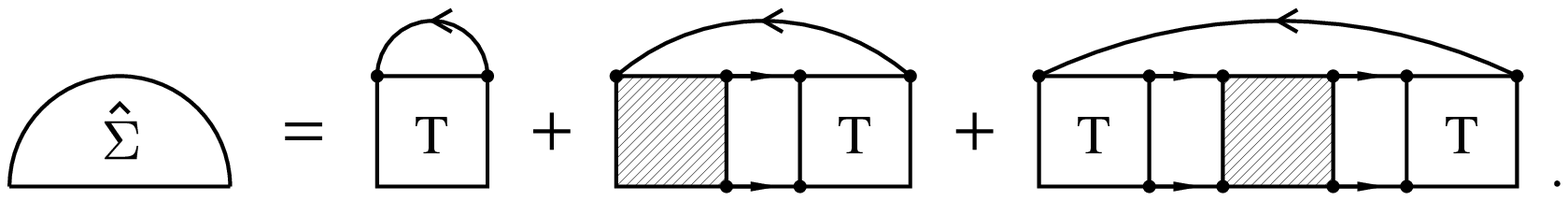,height=1.875cm,width=13.125cm}}
\end{eqnarray*}

Comparing these results with the predicted structure of the self-energies,
Eqs.~(\ref{Sigmastruc},\ref{Sigmainstruc}), the time-local contributions are
identified as
\begin{eqnarray}\label{Sigma-in_bin}
\Sigma^{IN}_{ac}&=&\pm i\hbar\, T_{ab,ef}G_{eg}G_{fh}C_{gh,cd}G_{db},\\
\label{Sigma_in_bin}
{\hat\Sigma}^{IN}_{ac}&=&\pm i\hbar\, C_{ab,ef}G_{eg}G_{fh}T_{gh,cd}G_{db},
\end{eqnarray}
or, diagrammatically,
\begin{eqnarray*}
\centerline{
\psfig{file=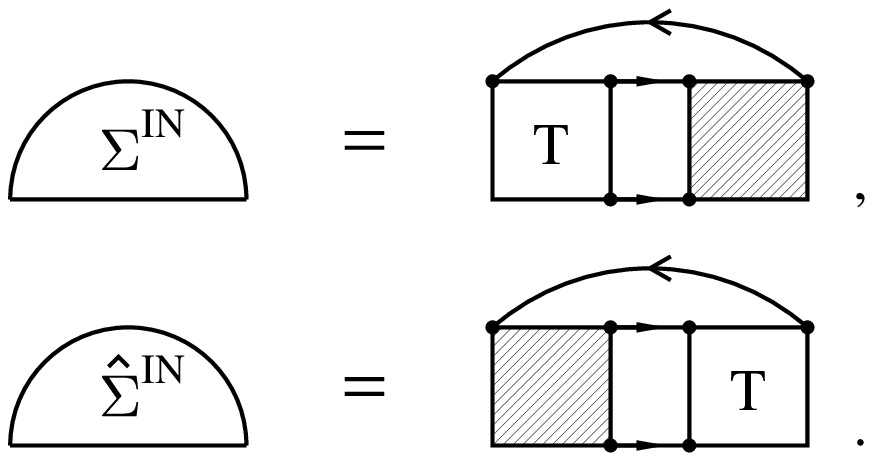,height=3.75cm,width=7.5cm}}
\end{eqnarray*}
Interestingly, the correlation part $\Sigma^C$ of the self-energy contains
an initial correlation contribution, too,
\begin{eqnarray}\label{Sigma-c_bin}
\Sigma^C_{ac}=\pm i\hbar\, T_{ab,cd}G_{db}
\pm (i\hbar)^2\, T_{ab,ef}G_{eg}G_{fh}C_{gh,ij}G_{ik}G_{jl}T_{kl,cd}G_{db},
\end{eqnarray}
and, again in diagrams,
\begin{eqnarray*}
\centerline{
\psfig{file=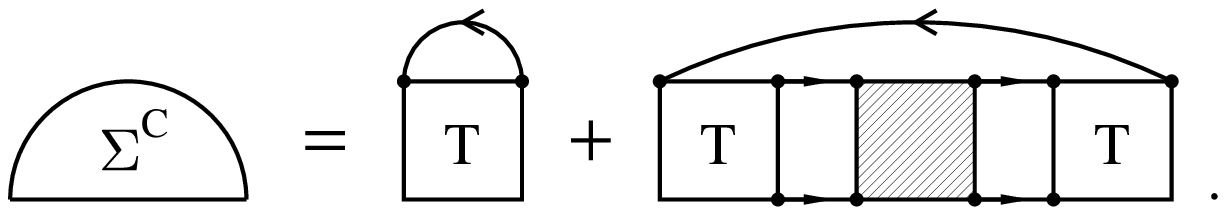,height=1.875cm,width=10.5cm}}
\end{eqnarray*}
With Eqs.~(\ref{Sigma_T_bin}--\ref{Sigma-c_bin}) we have found a
generalization of the T--matrix approximation. In addition to the usual ladder
term, the self-energies contain explicitly contributions of initial
correlations.

All relations derived so far are valid on the Keldysh--Schwinger contour.
In order to obtain the Kadanoff--Baym equations and kinetic equations for
the Wigner function, it is now necessary to specify the position of the time
arguments of Green's functions on the contour.
Then we obtain from the Dyson equation (\ref{Dyson}) the well-known
Kadanoff--Baym equations for the correlation functions $g^{\gtrless}$ (in the
following, small letters denote quantities on the physical time axis, and the
time arguments will be shown explicitly),
\begin{eqnarray}\label{KBG}
\int d{\bar t}\left\{s_{ac}(t,{\bar t})-\sigma^{HF}_{ac}(t,{\bar t})\right\}
g^{\gtrless}_{cb}({\bar t},t')
&=&\,\int\limits_{t_0}^t d{\bar t}
\left\{\sigma^>_{ac}(t,{\bar t})-\sigma^<_{ac}(t,{\bar t})\right\}
g^{\gtrless}_{cb}({\bar t},t')\nonumber\\
&&+\int\limits_{t_0}^{t'}\sigma^{\gtrless}_{ac}(t,{\bar t})
\left\{g^<_{cb}({\bar t},t')-g^>_{cb}({\bar t},t')\right\},\\
\label{KBGadj}
\int d{\bar t}\,g^{\gtrless}_{ac}(t,{\bar t})
\left\{s^{\dagger}_{cb}({\bar t},t')-\sigma^{HF}_{cb}({\bar t},t')\right\}
&=&\,\int\limits_{t_0}^t d{\bar t}
\left\{g^>_{ac}(t,{\bar t})-g^<_{ac}(t,{\bar t})\right\}
{\hat\sigma}^{\gtrless}_{cb}({\bar t},t')\nonumber\\
&&+\int\limits_{t_0}^{t'}g^{\gtrless}_{ac}(t,{\bar t})
\left\{{\hat\sigma}^<_{cb}({\bar t},t')-{\hat\sigma}^>_{cb}({\bar t},t')\right\}
.
\end{eqnarray}

The self-energies read in T--matrix approximation
\begin{eqnarray}\label{sigmat}
\sigma^{\gtrless}_{ac}(t,t')&=&\pm i\hbar\,t^{\gtrless}_{ab,cd}(t,t')
g^{\lessgtr}_{db}(t',t)
\pm i\hbar\,t^{C;IN}_{ab,cd}(t,t')g^{\lessgtr}_{db}(t',t)
\pm i\hbar\,t^{IN}_{ab,cd}(t,t')g^A_{db}(t_0,t),\\
\label{sigmadacht}
{\hat\sigma}^{\gtrless}_{ac}(t,t')&=&\pm i\hbar\,t^{\gtrless}_{ab,cd}(t,t')
g^{\lessgtr}_{db}(t',t)
\pm i\hbar\,t^{C;IN}_{ab,cd}(t,t')g^{\lessgtr}_{db}(t',t)
\pm i\hbar\,{\hat t}^{IN}_{ab,cd}(t,t')g^R_{db}(t',t_0),\\
\label{sigmaHF}
\sigma^{HF}_{ac}(t,t')&=&
\pm i\hbar\,\left(v_{ab,cd}\pm v_{ab,dc}\right)g^{\gtrless}_{db}(t,t')
\delta(t-t'),
\end{eqnarray}
where the initial correlation contributions are given by
\begin{eqnarray}\label{tin}
t^{IN}_{ab,cd}(t,t')&=&\int d{\bar t}\,t^R_{ab,ef}(t,{\bar t})
{\cal G}^R_{ef,gh}({\bar t},t_0)c_{gh,cd}(t_0)\delta(t_0-t'),\\
\label{tdachin}
{\hat t}^{IN}_{ab,cd}(t,t')&=&\int d{\bar t}\,c_{ab,ef}(t_0)
{\cal G}^A_{ef,gh}(t_0,{\bar t})t^A_{gh,cd}({\bar t},t')\delta(t_0-t),\\
\label{tcin}
t^{C;IN}_{ab,cd}(t,t')&=&i\hbar\int d{\bar t}\,d\overline{\overline{t}}\,
t^R_{ab,ef}(t,{\bar t})
{\cal G}^R_{ef,gh}({\bar t},t_0)c_{gh,ij}(t_0)
{\cal G}^A_{ij,kl}(t_0,\overline{\overline{t}})
t^A_{kl,cd}(\overline{\overline{t}},t'),
\end{eqnarray}
while the greater/less and the retarded/advanced T--Matrices obey the equations
\begin{eqnarray}\label{t><}
t^{\gtrless}_{ab,cd}(t,t')&=&
i\hbar\int d{\bar t}\,
v_{ab,ef}{\tilde{\cal G}}^R_{ef,gh}(t,{\bar t})t^{\gtrless}_{gh,cd}({\bar t},t')
+i\hbar\int d{\bar t}\,
v_{ab,ef}{\cal G}^{\gtrless}_{ef,gh}(t,{\bar t})t^A_{gh,cd}({\bar t},t'),\\
\label{tRA}
t^{R/A}_{ab,cd}(t,t')&=&v_{ab,cd}\delta(t-t')
+i\hbar\int d{\bar t}\,
v_{ab,ef}{\tilde{\cal G}}^{R/A}_{ef,gh}(t,{\bar t})t^{R/A}_{gh,cd}({\bar t},t'),
\end{eqnarray}
where we introduced the abbreviations
\begin{eqnarray}\label{calG}
{\cal G}^{R/A}_{ef,gh}(t,t')&=&g^{R/A}_{eg}(t,t')g^{R/A}_{fh}(t,t'),\qquad
{\cal G}^{\gtrless}_{ef,gh}(t,t')=g^{\gtrless}_{eg}(t,t')g^{\gtrless}_{fh}(t,t')
,\\\label{tildecalG}
{\tilde{\cal G}}^{R/A}_{ef,gh}(t,t')&=&\pm\Theta\left[\pm(t-t')\right]
\left\{{\cal G}^>_{ef,gh}(t,t')-{\cal G}^<_{ef,gh}(t,t')\right\}.
\end{eqnarray}
A further important relation is the optical theorem, which follows from
Eqs.~(\ref{t><}) and (\ref{tRA}),
\begin{eqnarray}\label{opt}
t^{\gtrless}_{ab,cd}(t,t')=i\hbar\int d{\bar t}\,d\overline{\overline{t}}\,
t^R_{ab,ef}(t,{\bar t})
{\cal G}^{\gtrless}_{ef,gh}({\bar t},\overline{\overline{t}})
t^A_{gh,cd}(\overline{\overline{t}},t').
\end{eqnarray}

Equations (\ref{KBG}--\ref{opt}) represent the Kadanoff--Baym equations in the
generalized binary collision approximation. Here, the T--matrix contains
contributions which are due to initial binary correlations. These
additional terms can be separated from the ``usual'' T--matrix, and, in
particular, do not influence the structure of the Lippmann--Schwinger equation
(\ref{tRA}).

\section{Non-Markovian Boltzmann equation}\label{Boltzmann}

In the previous section, we presented a far-reaching generalization of the
usual T--matrix approximation by incorporating initial correlations.
This way, the Kadanoff--Baym equations have become sufficiently general to
describe the evolution of a many-particle system on arbitrary time scales, in
particular on ultra-short times after an excitation. Their solutions, the
two-time correlation functions, contain a tremendous amount of information on
the statistical and dynamical properties of a strongly correlated many-particle
system, fully including damping (lifetime) of the one and two-particle states
\cite{note3}. However, in many cases the information contained in the Wigner
distribution is sufficient. Therefore, in the following, we will derive an
equation for this function, i.e., a kinetic equation in a narrow sense.

For this purpose, we consider the Kadanoff--Baym equations
(\ref{KBG},\ref{KBGadj}) in the limit of equal times $t=t'$ and subtract them
from each other. The result is an equation for the distribution function
which reads, in momentum representation (we consider a homogeneous system
without external forces),
\begin{eqnarray}\label{ZDG}
\frac{\partial}{\partial t}f({\bf p},t)&=&
\pm\int\limits_{t_0}^t d{\bar t}\,
\left\{\sigma^>({\bf p},t,{\bar t})g^<({\bf p},{\bar t},t)
-\sigma^<({\bf p},t,{\bar t})g^>({\bf p},{\bar t},t)\right.\nonumber\\
&&\hspace{7ex}\left.+g^<({\bf p},t,{\bar t}){\hat\sigma}^>({\bf p},{\bar t},t)
-g^>({\bf p},t,{\bar t}){\hat\sigma}^<({\bf p},{\bar t},t)\right\}\nonumber\\
&=&I({\bf p},t)+I^{IC}({\bf p},t).
\end{eqnarray}
This so-called time-diagonal equation is a very general representation of a
kinetic equation. The r.h.s. describes the influence of collisions as well as
initial correlations on the Wigner distribution and is, in principle,
determined by the exact self-energy and the two-time correlation functions.

In order to obtain a closed kinetic equation, two major tasks remain:
(i) an approximation for the self-energies has to be chosen, and (ii)
the reconstruction problem, i.e., the determination of $g^{\gtrless}$ as a
functional of the Wigner distribution, has to be solved.
The first task has already been dealt with in the previous section, with the
result being the generalized T--matrix approximation, given by
Eqs.~(\ref{sigmat}--\ref{opt}).
Let us now consider the reconstruction problem. In order to obtain the
functional relation $g^{\gtrless}=g^{\gtrless}[f]$, we use the generalized
Kadanoff--Baym ansatz (GKBA) proposed by Lipavsk\'{y} et al.
\cite{lipavski-etal.86},
\begin{eqnarray}\label{GKBA}
g^{\gtrless}({\bf p},t,t')=\pm\left\{g^R({\bf p},t,t')f^{\gtrless}({\bf p},t')
-f^{\gtrless}({\bf p},t)g^A({\bf p},t,t')\right\},
\end{eqnarray}
with $f^<=f$ and $f^>=1\pm f$. For the products ${\cal G}^{\gtrless}$ then
follows
\begin{eqnarray}\label{ggGKBA}
{\cal G}^{\gtrless}_{12}(t,t')=
{\cal G}^R_{12}(t,t')F^{\gtrless}_{12}(t')
+F^{\gtrless}_{12}(t){\cal G}^A_{12}(t,t'),
\end{eqnarray}
where we used the abbreviations
$F^{\gtrless}_{12}=f^{\gtrless}({\bf p}_1)f^{\gtrless}({\bf p}_2)$ and
${\cal G}_{12}={\cal G}({\bf p}_1,{\bf p}_2)$.
From Eq.~(\ref{ggGKBA}) follow relations between the functions
${\cal G}^{R/A}$ and ${\tilde{\cal G}}^{R/A}$ which were defined in
Eqs.~(\ref{calG},\ref{tildecalG}),
\begin{eqnarray}\label{G-G_R}
{\tilde{\cal G}}^R_{12}(t,t')&=&{\cal G}^R_{12}(t,t')N_{12}(t'),\\
{\tilde{\cal G}}^A_{12}(t,t')&=&-N_{12}(t){\cal G}^A_{12}(t,t'),
\end{eqnarray}
where we introduced the Pauli blocking factor
$N_{12}=1\pm f({\bf p}_1)\pm f({\bf p}_2)$.

Now we insert the self-energies in T--matrix approximation, Eqs.
(\ref{sigmat}--\ref{tcin}), into the time diagonal equation (\ref{ZDG}),
replacing $t^{\gtrless}$ with the help of the optical theorem (\ref{opt})
and ${\cal G}^{\gtrless}$ by means of the reconstruction ansatz (\ref{ggGKBA}).
The result is the collision integral $I$,
\begin{eqnarray}\label{I}
&&I({\bf p}_1,t)=(i\hbar)^2\int \frac{d{\bf p}_2}{(2\pi\hbar)^3}
\frac{d{\bar{\bf p}}_1}{(2\pi\hbar)^3}\frac{d{\bar{\bf p}}_2}{(2\pi\hbar)^3}
\int d{\bar t}\,d\overline{\overline{t}}\,d\overline{\overline{\overline{t}}}
\nonumber\\
&\times&\left\{
t^R({\bf p}_1{\bf p}_2\,t,{\bar{\bf p}}_1{\bar{\bf p}}_2\,{\bar t})
{\bar{\cal G}}_{12}^R({\bar t},\overline{\overline{t}})
t^A({\bar{\bf p}}_1{\bar{\bf p}}_2\,\overline{\overline{t}},
{\bf p}_1{\bf p}_2\,\overline{\overline{\overline{t}}})
{\cal G}_{12}^A(\overline{\overline{\overline{t}}},t)
\left[{\bar F}_{12}^>(\overline{\overline{t}})
F_{12}^<(\overline{\overline{\overline{t}}})
-{\bar F}_{12}^<(\overline{\overline{t}})
F_{12}^>(\overline{\overline{\overline{t}}})\right]\right.
\nonumber\\
&&+t^R({\bf p}_1{\bf p}_2\,t,{\bar{\bf p}}_1{\bar{\bf p}}_2\,{\bar t})
{\bar{\cal G}}_{12}^A({\bar t},\overline{\overline{t}})
t^A({\bar{\bf p}}_1{\bar{\bf p}}_2\,\overline{\overline{t}},
{\bf p}_1{\bf p}_2\,\overline{\overline{\overline{t}}})
{\cal G}_{12}^A(\overline{\overline{\overline{t}}},t)
\left[{\bar F}_{12}^>({\bar t})
F_{12}^<(\overline{\overline{\overline{t}}})
-{\bar F}_{12}^<({\bar t})
F_{12}^>(\overline{\overline{\overline{t}}})\right]
\nonumber\\
&&-\,{\cal G}_{12}^R(t,{\bar t})
t^R({\bf p}_1{\bf p}_2\,{\bar t},
{\bar{\bf p}}_1{\bar{\bf p}}_2\,\overline{\overline{t}})
{\bar{\cal G}}_{12}^A
(\overline{\overline{t}},\overline{\overline{\overline{t}}})
t^A({\bar{\bf p}}_1{\bar{\bf p}}_2\,\overline{\overline{\overline{t}}},
{\bf p}_1{\bf p}_2\,t)
\left[F_{12}^>({\bar t})
{\bar F}_{12}^<(\overline{\overline{t}})
-F_{12}^<({\bar t})
{\bar F}_{12}^>(\overline{\overline{t}})\right]
\nonumber\\
&&-\,{\cal G}_{12}^R(t,{\bar t})
t^R({\bf p}_1{\bf p}_2\,{\bar t},
{\bar{\bf p}}_1{\bar{\bf p}}_2\,\overline{\overline{t}})
{\bar{\cal G}}_{12}^R
(\overline{\overline{t}},\overline{\overline{\overline{t}}})
t^A({\bar{\bf p}}_1{\bar{\bf p}}_2\,\overline{\overline{\overline{t}}},
{\bf p}_1{\bf p}_2\,t)
\left.\left[F_{12}^>({\bar t})
{\bar F}_{12}^<(\overline{\overline{\overline{t}}})
-F_{12}^<({\bar t})
{\bar F}_{12}^>(\overline{\overline{\overline{t}}})\right]
\right\},
\end{eqnarray}
with
${\bar{\cal G}}_{12}^{R/A}={\cal G}^{R/A}({\bar{\bf p}}_1,{\bar{\bf p}}_2)$
and
$t^{R/A}({\bf p}_1{\bf p}_2\,t,
{\bar{\bf p}}_1{\bar{\bf p}}_2\,{\bar t})=
\left\langle {\bf p}_1{\bf p}_2\left|t^{R/A}(t,{\bar t})\right|
{\bar{\bf p}}_1{\bar{\bf p}}_2\right\rangle$,
and the collision integral arising from initial correlations $I^{IC}$,
\begin{eqnarray}\label{I-IC}
I^{IC}({\bf p}_1,t)&=&i\hbar\int \frac{d{\bf p}_2}{(2\pi\hbar)^3}
\frac{d{\bar{\bf p}}_1}{(2\pi\hbar)^3}
\frac{d{\bar{\bf p}}_2}{(2\pi\hbar)^3}
\int d{\bar t}\nonumber\\
&&\times\left\{
t^R({\bf p}_1{\bf p}_2\,t,{\bar{\bf p}}_1{\bar{\bf p}}_2\,{\bar t})
{\cal K}({\bar{\bf p}}_1{\bar{\bf p}}_2\,{\bar t},{\bf p}_1{\bf p}_2\,t)
-{\cal K}({\bf p}_1{\bf p}_2\,t,{\bar{\bf p}}_1{\bar{\bf p}}_2\,{\bar t})
t^A({\bar{\bf p}}_1{\bar{\bf p}}_2\,{\bar t},{\bf p}_1{\bf p}_2\,t)
\right\}\nonumber\\
&-&(i\hbar)^2\int \frac{d{\bf p}_2}{(2\pi\hbar)^3}
\frac{d{\bar{\bf p}}_1}{(2\pi\hbar)^3}\frac{d{\bar{\bf p}}_2}{(2\pi\hbar)^3}
\frac{d\overline{\overline{{\bf p}}}_1}{(2\pi\hbar)^3}
\frac{d\overline{\overline{{\bf p}}}_2}{(2\pi\hbar)^3}
\int d{\bar t}\,d\overline{\overline{t}}\,d\overline{\overline{\overline{t}}}
\nonumber\\
&&\times\left\{
t^R({\bf p}_1{\bf p}_2\,t,{\bar{\bf p}}_1{\bar{\bf p}}_2\,{\bar t})
{\cal K}({\bar{\bf p}}_1{\bar{\bf p}}_2\,{\bar t},
\overline{\overline{{\bf p}}}_1\overline{\overline{{\bf p}}}_2\,
\overline{\overline{t}})
t^A(\overline{\overline{{\bf p}}}_1\overline{\overline{{\bf p}}}_2\,
\overline{\overline{t}},
{\bf p}_1{\bf p}_2\,\overline{\overline{\overline{t}}})
N_{12}(\overline{\overline{\overline{t}}})
{\cal G}_{12}^A(\overline{\overline{\overline{t}}},t)\right.
\nonumber\\
&&\quad \left.+{\cal G}_{12}^R(t,{\bar t})N_{12}({\bar t})
t^R({\bf p}_1{\bf p}_2\,{\bar t},
{\bar{\bf p}}_1{\bar{\bf p}}_2\,\overline{\overline{t}})
{\cal K}({\bar{\bf p}}_1{\bar{\bf p}}_2\,\overline{\overline{t}},
\overline{\overline{{\bf p}}}_1\overline{\overline{{\bf p}}}_2\,
\overline{\overline{\overline{t}}})
t^A(\overline{\overline{{\bf p}}}_1\overline{\overline{{\bf p}}}_2\,
\overline{\overline{\overline{t}}},{\bf p}_1{\bf p}_2\,t)
\right\},\\
\mbox{with}\quad
&&{\cal K}({\bf p}_1{\bf p}_2\,t,{\bar{\bf p}}_1{\bar{\bf p}}_2\,{\bar t})=
{\cal G}_{12}^R(t,t_0)
c({\bf p}_1{\bf p}_2,{\bar{\bf p}}_1{\bar{\bf p}}_2;t_0)
{\bar{\cal G}}_{12}^A(t_0,{\bar t}).
\end{eqnarray}

With Eqs.~(\ref{ZDG},\ref{I},\ref{I-IC}) we have obtained a very general
quantum kinetic equation. The character of its approximations goes far beyond
that of the usual Boltzmann equation. The collision integral $I({\bf p}_1,t)$
was derived without any approximation with respect to the times and thus fully
includes retardation and memory effects which is usually referred to as
non-Markovian behavior. Many-particle effects, as for instance self-energy and
damping \cite{note3}, and spin statistics effects (Pauli blocking) are included.
So far, no restriction has been introduced with respect to the retarded and
advanced propagators ${\cal G}^{R/A}$.
In principle, they are to be determined self-consistently from their KBE which
follow easily from Eq.~(\ref{Dyson}). However, to avoid this essential
complication, in most cases approximations are used. For example, in the
quasiparticle approximation, the propagators are given explicitly by
\begin{eqnarray}\label{GRAqp}
{\cal G}_{12}^{R/A}(t,t')=\frac{1}{(i\hbar)^2}\Theta[\pm(t-t')]
e^{\frac{i}{\hbar}[E_{12}+i\Gamma_{12}](t-t')}
\end{eqnarray}
with $E_{12}=\frac{p_1^2}{2m}+\frac{p_2^2}{2m}
+{\rm Re}\,\sigma_1^R+{\rm Re}\,\sigma_2^R$ and
$\Gamma_{12}={\rm Im}\,\sigma_1^R+{\rm Im}\,\sigma_2^R$.

Furthermore, the retarded and advanced T--Matrices are many-particle
generalizations of the familiar T--Matrices of quantum scattering theory. They
have to be determined from the Lippmann--Schwinger equation (\ref{tRA}) which
reads in momentum representation
\begin{eqnarray}\label{tRA-p}
t^{R/A}({\bf p}_1{\bf p}_2\,t,{\bf p}'_1{\bf p}'_2\,t')&=&
v({\bf p}_1-{\bf p}'_1)(2\pi\hbar)^3
\delta({\bf p}_1+{\bf p}_2-{\bf p}'_1-{\bf p}'_2)
\delta(t-t')\nonumber\\
&&+i\hbar\int\frac{d{\bar{\bf p}}_1}{(2\pi\hbar)^3}
\frac{d{\bar{\bf p}}_2}{(2\pi\hbar)^3}
\int d{\bar t}\,
v({\bf p}_1-{\bar{\bf p}}_1)(2\pi\hbar)^3
\delta({\bf p}_1+{\bf p}_2-{\bar{\bf p}}_1-{\bar{\bf p}}_2)
\nonumber\\
&&\;\times{\tilde{\cal G}}^{R/A}({\bar{\bf p}}_1,{\bar{\bf p}}_2;t,{\bar t})
t^{R/A}({\bar{\bf p}}_1{\bar{\bf p}}_2\,{\bar t},{\bf p}'_1{\bf p}'_2\,t').
\end{eqnarray}

The collision integral $I^{IC}({\bf p}_1,t)$ contains the terms arising from
binary correlations, existing in the system initially. It should be stressed
explicitly that the structure of these contributions is completely general and
does not depend on parameters characterizing the system, such as coupling
strength or degree of degeneracy. Furthermore, the inclusion of initial
correlations does not depend on their actual form, i.e. the form of the
function $c$. The damping of the two-particle propagators leads to a decay of
this collision term, i.e. the initial correlations die out on a time scale
which is determined by the one-particle damping rates \cite{note3}.

Finally, we want to remark here that our result for the non-Markovian Boltzmann
equation is in agreement with the result derived within the framework of
the density operator technique, see \cite {KBKS97}.

\section*{Acknowledgments}

The authors acknowledge discussions with Th.~Bornath (Rostock).
This work is supported by the Deutsche Forschungsgemeinschaft (SFB 198 and
Schwerpunkt ``Quantenkoh\"arenz in Halbleitern'').

\appendix
\section{Solution of the generalized T--matrix equations}

We rewrite Eqs.~(\ref{T_Leiter},\ref{Phi_Leiter}), which contain the
ladder-type terms of the generalized T--matrix (\ref{Tdef}),
\begin{eqnarray}\label{T_Leiter_A}
{\cal T}_{ab,cd}&=&V_{ab,cd}+\Phi_{ab,cd}+V_{ab,ef}G_{eg}G_{fh}C_{gh,cd}
+i\hbar\,V_{ab,ef}G_{eg}G_{fh}{\cal T}_{gh,cd},\\
\label{Phi_Leiter_A}
\Phi_{ab,cd}&=&C_{ab,ef}G_{eg}G_{fh}V_{gh,cd}
+i\hbar\,\Phi_{ab,ef}G_{eg}G_{fh}V_{gh,cd}.
\end{eqnarray}

Due to the structure of Eq.~(\ref{T_Leiter_A}), ${\cal T}$ can be split into
three parts,
\begin{eqnarray}\label{T3}
{\cal T}_{ab,cd}&=&
{\cal T}^{(A)}_{ab,cd}+{\cal T}^{(B)}_{ab,cd}+{\cal T}^{(C)}_{ab,cd},\\
\label{TA}
{\cal T}^{(A)}_{ab,cd}&=&
V_{ab,cd}+i\hbar\,V_{ab,ef}G_{eg}G_{fh}{\cal T}^{(A)}_{gh,cd},\\
\label{TB}
{\cal T}^{(B)}_{ab,cd}&=&V_{ab,ef}G_{eg}G_{fh}C_{gh,cd}
+i\hbar\,V_{ab,ef}G_{eg}G_{fh}{\cal T}^{(B)}_{gh,cd},\\
\label{TC}
{\cal T}^{(C)}_{ab,cd}&=&
\Phi_{ab,cd}+i\hbar\,V_{ab,ef}G_{eg}G_{fh}{\cal T}^{(C)}_{gh,cd}.
\end{eqnarray}
Obviously, Eq.~(\ref{TA}) coincides with the well-known ladder equation of the
T--matrix approximation. Thus, we can identify ${\cal T}^{(A)}$ with the
usual T--matrix $T$. The ladder equation (\ref{TA}) now serves as a basis for
the solution of (\ref{TB}) and (\ref{TC}). If one assumes for ${\cal T}^{(B)}$
the form
\begin{eqnarray}\label{TB_lsg}
{\cal T}^{(B)}_{ab,cd}=T_{ab,ef}G_{eg}G_{fh}C_{gh,cd},
\end{eqnarray}
Eq.~(\ref{TB}) is valid if (\ref{TA}) holds.
In order to determine ${\cal T}^{(C)}$, Eq.~(\ref{Phi_Leiter_A}) has to be
considered. This equation is fulfilled if $\Phi$ is of the structure
\begin{eqnarray}\label{Phi_lsg}
\Phi_{ab,cd}=C_{ab,ef}G_{eg}G_{fh}T_{gh,cd},
\end{eqnarray}
if the adjoint equation to (\ref{TA}) is valid. Due to the symmetry properties
of $T$, Eq.~(\ref{TA}) and its adjoint are equivalent.
Inserting (\ref{Phi_lsg}) into Eq.~(\ref{TC}) and assuming ${\cal T}^{(C)}$ to
be of the structure
\begin{eqnarray}\label{Tc_lsg}
{\cal T}^{(C)}_{ab,cd}=C_{ab,ef}G_{eg}G_{fh}T_{gh,cd}
+i\hbar\,T_{ab,ef}G_{eg}G_{fh}C_{gh,ij}G_{ik}G_{jl}T_{kl,cd},
\end{eqnarray}
Eq.~(\ref{TC}) is fulfilled, again under the assumption (\ref{TA}).
Collecting all parts together, ${\cal T}$ can be represented as
\begin{eqnarray}\label{T_expl_A}
{\cal T}_{ab,cd}=T_{ab,cd}
&+&i\hbar\, T_{ab,ef}G_{eg}G_{fh}C_{gh,ij}G_{ik}G_{jl}T_{kl,cd}
\nonumber\\
&+&T_{ab,ef}G_{eg}G_{fh}C_{gh,cd}
+C_{ab,ef}G_{eg}G_{fh}T_{gh,cd},
\end{eqnarray}
together with the equation for the well-known ``ladder T--matrix'',
\begin{eqnarray}\label{T_LSG_A}
T_{ab,cd}=V_{ab,cd}+i\hbar\,V_{ab,ef}G_{eg}G_{fh}T_{gh,cd}.
\end{eqnarray}


\begin{thebibliography}{10}
\bibitem{note1} In order to present the equations in a compact way, we will use
a representationless index notation with summation convention \cite{BW63}.
The indices include the time arguments, too, unless they are written explicitly.
\bibitem{note2} In \cite{SKB99}, ${\hat\Sigma}^{IN}$ in the argument of the
functional derivative on the r.h.s. of (\ref{Sigma}) is not written explicitly.
However, it should be mentioned here, that there is a coupling between the
equation for $\Sigma$ and its adjoint.
\bibitem{note3} We should mention here that, in the frame of the T--matrix
approximation with bare interaction, the damping of the two-particle
states is included on the one-particle level. A real two-particle damping would
require to include into the T--matrix equation three-particle interactions, or
the dynamical screening, respectively.
\end{thebibliography}
\end{document}